\documentclass[prc,amsmath,twocolumn,showkeys,superscriptaddress]{revtex4}
\usepackage{gensymb}
\usepackage{graphicx,color}
\usepackage{amssymb}
\usepackage{enumerate}
\usepackage{verbatim}
\usepackage{natbib}
\usepackage{float}
\usepackage[outdir=./]{epstopdf}

\begin{document}
  \newcommand {\nc} {\newcommand}
  \nc {\Sec} [1] {Sec.~\ref{#1}}
  \nc {\IBL} [1] {\textcolor{black}{#1}} 
  \nc {\IR} [1] {\textcolor{red}{#1}} 
  \nc {\IB} [1] {\textcolor{blue}{#1}} 
  \nc {\IG} [1] {\textcolor{green}{#1}}

\title{Exploring experimental conditions to reduce uncertainties in the optical potential}

\author{M. Catacora-Rios}
\affiliation{National Superconducting Cyclotron Laboratory, Michigan State University, East Lansing, MI 48824}
\affiliation{Department of Physics and Astronomy, Michigan State University, East Lansing, MI 48824-1321}
\author{G.~B.~King}
\affiliation{National Superconducting Cyclotron Laboratory, Michigan State University, East Lansing, MI 48824}
\affiliation{Department of Physics and Astronomy, Michigan State University, East Lansing, MI 48824-1321}
\affiliation{Department of Physics, Washington University, St. Louis, MO 63130, USA}
\author{A.~E.~Lovell} 
\affiliation{Theoretical Division, Los Alamos National Laboratory, Los Alamos, NM 87545}
\affiliation{Center for Nonlinear Studies, Los Alamos National Laboratory, Los Alamos, NM 87545}
\author{F.~M.~Nunes}
\email{nunes@nscl.msu.edu}
\affiliation{National Superconducting Cyclotron Laboratory, Michigan State University, East Lansing, MI 48824}
\affiliation{Department of Physics and Astronomy, Michigan State University, East Lansing, MI 48824-1321}

\date{\today}


\begin{abstract}
\begin{description}
\item[Background:] Uncertainty quantification for nuclear theories has gained a more prominent role in the field, with more and more groups attempting to understand the uncertainties on their calculations. However, recent studies have shown that the uncertainties on the optical potentials are too large for the theory to be useful. 
\item[Purpose:] The purpose of this work is to explore possible experimental conditions that may reduced the uncertainties on elastic scattering and single-nucleon transfer cross sections that come from the fitting of the optical model parameters to experimental data.
\item[Method:] Using Bayesian methods, we explore the effect of the uncertainties of optical model parameters on the angular grid of the differential cross section, including cross section data at nearby energies, and changes in the experimental error bars.  We also study the effect on the resulting uncertainty when other observables are included in the fitting procedure, particularly the total (reaction) cross sections.
\item[Results:] We study proton and neutron elastic scattering on $^{48}$Ca and $^{208}$Pb. We explore the parameter space with Markov-Chain Monte Carlo, produce posterior distributions for the optical model parameters, and construct the corresponding 95\% confidence intervals on the elastic-scattering cross sections. We also propagate the uncertainties on the optical potentials to the $^{48}$Ca(d,p)$^{49}$Ca(g.s.) and $^{208}$Pb(d,p)$^{209}$Pb(g.s.) cross sections.
\item[Conclusions:]  We find little sensitivity to the angular grid and an improvement of up to a factor of 2 on the uncertainties by including data at a nearby energy. Although reducing the error bars on the data does reduce the uncertainty, the gain is often considerably smaller than one would naively expect. We also find that the inclusion of total reaction cross section can improve the uncertainty although the magnitude of the effect depends strongly on the cases considered. 
\end{description}
\end{abstract}


\maketitle

\section{Introduction}
\label{intro}

Nuclear reactions offer  useful and versatile probes in the study of nuclear structure and astrophysics. For example, nucleon elastic scattering provides information on the effective interactions between projectile and target, and single-nucleon (d,p) transfer enables the study of the single-particle configuration of orbitals in the final nucleus. \IBL{On the astrophysics side, we know that a good fraction of the heavy nuclei were generated through neutron capture reactions on unstable isotopes. Here too, (d,p) reactions offer an important indirect probe since direct neutron capture measurements are not feasible. No matter the application, for a meaningful interpretation of nuclear reaction data, one needs reliable reaction theory.}

\IBL{When intermediate and heavy nuclei are involved, most often treating  the reaction process in a fully microscopic ab-initio approach is not tractable. Instead, few-body theories for reactions are developed, having as input the effective interactions between the composite particles (the so-called  optical potentials). }

\IBL{Over the decades there has been much work toward developing nucleon optical potentials (e.g. \cite{Feshbach_ap1962,Jeukenne_prc1977,Bauge_prc1998,Bauge_prc2001}) however, more recently, the focus is increasingly moving toward extracting this quantity from first principles, without any fitting parameters (e.g. \cite{Rotureau_prc2017,rotureau2018,burrows2018,gennari2018,idini2019,whitehead2019}). For lighter systems, computations can cover most of Hilbert space and predictions compare well with data (e.g. \cite{hupin2014}). For intermediate mass systems, the ab-initio approaches face challenges in estimating correctly the absorption to other channels, due to truncations in the model space. For heavy systems,  ab-initio approaches are just not feasible. While these fully microscopic efforts are important and should be pursued, it is clear that semi-microscopic approaches, such as the dispersive optical model \cite{charity2007,dom-local,Mahzoon_prl2014},  are currently more promising: they provide a good description of the data while retaining some physical insight based on theory. Nevertheless, for studying effects across the nuclear chart, often the only  alternative is to use modern global phenomenological optical potentials, which span a range of energies and masses. These are obtained from fitting a large body of data including elastic scattering, total cross sections, and often analyzing powers on most stable nuclei  (e.g. \cite{bg69,kd2003,ch89}). }

\IBL{Just as important as developing reliable optical potentials for reactions, is the quantification of the uncertainties associated with these potentials. The primary objective of this work is not  obtaining new optical potentials but rather, through state-of-the-art statistical techniques, understanding and quantifying their uncertainties and developing a path toward reducing them systematically.}

Over the past several years, the rigorous quantification of theoretical uncertainties in low energy nuclear physics has been gaining traction, from Effective Field Theory \cite{Furnstahl2015,Perez2015,Wesolowski2016,Schindler2009} to Density Functional Theory \cite{Schunck2015,McDonnell2015} and from \emph{ab initio} methods \cite{Stroberg2019} to few-body reaction models \cite{Lovell2015,Lovell2017,Lovell2018,King2018,King2019}.  The focus has also recently shifted from the propagation of uncertainties using covariance matrices, denoted here as frequentist methods and defined by a $\chi^2$ minimization, to more sophisticated Bayesian methods which provide a pathway for quantifying both parametric and model uncertainties.  Most recently, DFT model comparisons are being made using methods trending toward machine learning techniques, with Gaussian Processes and Bayesian Neural Networks \cite{Neufcourt2018,Neufcourt2019}.

\IBL{The present work comes in the sequence of a number of uncertainty quantification (UQ) studies \cite{Lovell2015,Lovell2017,Lovell2018,King2018,King2019}: the goal is to use modern statistical tools to reliably understand, quantify and control uncertainties in the theory for direct reactions.} 
Over the last few years, our UQ efforts have focused on the parametric uncertainties associated with the nucleon-target optical potential, when informed by elastic scattering, and understanding how those uncertainties  propagate to deuteron induced transfer reactions.  
Beginning with standard covariance propagation methods \cite{Lovell2017,King2018} and moving on to Bayesian \cite{Lovell2018}, we have quantified uncertainties from the fitting of optical potentials to nucleon elastic scattering and then propagated them to transfer cross sections, using both the Distorted Wave Born Approximation (DWBA) and the Adiabatic Wave Approximation (ADWA).  We have also made a direct and systematic comparison between the frequentist $\chi^2$ optimization and Bayesian methods \cite{King2019}.  This study showed that, despite popular belief, the two methods are not identical and that, for the higher levels of confidence, frequentist methods severely underestimate the uncertainties while the Bayesian approach provides a truer representation of the uncertainty.  Overall, uncertainties on transfer cross sections obtained from the Bayesian approach ranged from 40\% to over 100\%. Such large uncertainties render these probes less useful for extracting structure or astrophysical information. It is desirable that the parametric uncertainties do not exceed the errors on the experimental data, which are typically of the order of 10\%.  

When the model relies on well defined expansions, the reduction of uncertainties  is, in principle, straight-forward.  This is the case for effective field theories: due to the order-by-order nature of the problem, the uncertainties can be reduced by adding each successive order \cite{Furnstahl2015a,Melendez2017}.  The complexity in uncertainty quantification increases for models that are not expressed as expansions. This is the case for DFT calculations: the source of uncertainties can come from both the imprecise form of the functionals and the specific choice of the data protocol used to optimize the functionals \cite{Dobaczewski2014}.  Due to the non-perturbative nature of the reactions we are interested in, the few-body model used to describe the reaction does not offer an order-by-order systematic improvement on the uncertainties. Like DFT, improving uncertainties in the optical potential will most likely come from including more data into the fitting procedure. 

\IBL{The goal of the present work is to explore different avenues to reduce the uncertainties found in \cite{Lovell2017,Lovell2018,King2018,King2019}.
The UQ methods we begin to explore here fall under the umbrella of Bayesian experimental design \cite{exp-design}  and should be applicable across the nuclear chart, whenever the concept of an optical potential holds. For this reason, and to optimize computations, we use phenomenological potentials in the current study, but underline that the UQ tools developed are general and can  be coupled with other optical potential frameworks.}
  
In this study, we investigate four different aspects of the data with the intent to reduce the uncertainties coming from the optical potential.  We explore: i) different ranges for the angles at which scattering is measured, ii) the use of nearby beam energies, iii) the magnitude of the experimental errors, and iv) the addition of reaction data beyond differential elastic cross sections.  Here, we present applications to reactions on  $^{48}$Ca and $^{208}$Pb. 

This paper is organized in the following way.  In Sec. \ref{theory}, we briefly discuss the theoretical models used in the current work.  Section \ref{results} contains results and a discussion.  Finally, conclusions are drawn in Sec. \ref{conclusions}.

\section{Theoretical considerations and inputs}
\label{theory}

\subsection{Bayesian Statistical Framework}

In Bayes' statistics, one tests a hypothesis $H$ (model)  against some constraining external information $D$ (data). Bayes' theorem tells us that:

\begin{equation}
p(H|D) = \frac{p(H)p(D|H)}{p(D)},
\label{bayes}
\end{equation}
where $p(H|D)$ is the posterior distribution of the hypothesis, conditional on the data and $p(H)$ represents prior information. 

In our work \cite{Lovell2018,King2019}, the hypothesis  is the optical model with parameters $x_j$ ($j=1,N_{par}$) and the data are the elastic scattering angular distributions $\sigma(\theta_i)$ (with $i=1,N_{\theta}$). The prior distributions $p(H)$, summarize our knowledge before the data are seen, and the likelihood function, $p(D|H)$, contains information about how well the model reproduces the data. Typically, as in \cite{Lovell2018}, we use a standard normal distribution for the likelihood, ${\cal L}=e^{-\chi^2/2}$. When considering only elastic scattering angular distributions, $\chi^2$ becomes:
\begin{equation}
\chi^2 = \frac{1}{N_{\theta}} \sum_{i=1}^{N_{\theta}} \frac{(\sigma^{th}(\theta_i) - \sigma^{exp}(\theta_i))^2}{\Delta\sigma_i^2},
\end{equation}
with $\sigma(\theta_i)$ being the elastic angular distribution at a given angle $\theta_i$ and $\Delta \sigma _i$ the experimental uncertainties at $\theta_i$. 

The remaining piece in Eq. \ref{bayes} is the evidence, $p(D)$. Evaluating the Bayesian evidence is numerically difficult and often intractable, and therefore Monte Carlo methods are needed to sample the posterior distribution of parameters. Here, we use the Metropolis-Hastings Markov Chain Monte Carlo (numerical details can be found in \cite{Lovell2018}). 

Once we have the posterior distributions of optical model parameters from the elastic scattering Bayesian fitting, we can use these to generate predictions for the ADWA transfer cross sections.  

\subsection{Optical Potential and elastic scattering}

Optical potentials, $U_{opt}$, capture the complex many-body effects of nucleon-nucleus scattering. These potentials contain a real part, representing the mean field seen by the nucleon approaching the target, and an imaginary part that accounts for flux that leaves the elastic channel. 
In general, optical potentials contain: i) a real volume term of Woods-Saxon shape with parameters $V$, $r$, and $a$ for the depth, radius, and diffuseness; ii) an imaginary volume term (of Woods-Saxon shape) with parameters $W_{v}$, $r_{v}$, and $a_{v}$; iii) a surface imaginary term (derivative of Woods-Saxon shape) with parameters $W_{s}$, $r_{s}$, and $a_{s}$; iv) a standard spin-orbit term; and v) a regular Coulomb term for charged projectiles. In this work, the spin-orbit and Coulomb terms are kept constant, but all other terms are allowed to vary. Thus, we typically deal with 9 parameters.

As in previous work,  optical potential parameters are initialized using the Becchetti and Greenlees (BG) global parameterization \cite{bg69}. 
In order to avoid restricting ourselves to the limited available data, we use mock data generated from the Koning-Delaroche (KD) global optical potential \cite{kd2003}. This allows us total freedom in exploring angular and energy discretization of observables. Unless otherwise stated, we take the error on the data to be 10\%.
Concerning the Bayesian method,
wide Gaussian priors, centered on the original BG parameter value and with a standard deviation equal to the mean value of the distribution, were chosen to ensure that parameter space was adequately sampled. In some cases, one of the imaginary depths of the BG parameterization can be zero; when this occurs, we take the parameter value to be 1 MeV with a width of 10 MeV to adequately sample this piece of the potential as well. Then, 1600 parameter sets were drawn to create 95\% confidence intervals by taking the densest 95\% of the cross section values at each angle. The wrapper codes used to perform these calculations make use of the reaction codes {\sc fresco} and {\sc sfresco} \cite{fresco}.

\subsection{Transfer cross sections}

Following the quantification of uncertainties in nucleon elastic scattering, we also investigate how these uncertainties propagate to single-nucleon (d,p) transfer reactions.
The model here used to describe (d,p) reactions  is the adiabatic wave approximation (ADWA) \cite{Johnson1974}, which provides an effective simple way of incorporating deuteron breakup to all orders in the reaction formalism. In this formalism, one starts from a three-body Hamiltonian of the $n + p + A$ system, and the key inputs are the pairwise interactions, namely proton-target and neutron-target optical potentials, and the well known proton-neutron interaction. In ADWA, the cross section can be directly obtained from the following T-matrix:
\begin{equation}
T = \langle \phi_{nA}\chi^{(-)}_{pB}|V_{np}|\phi_{np}\chi^{ad}_{d} \rangle,
\end{equation}
where the adiabatic wave $\chi^{ad}_d$ is generated from the effective adiabatic potential:
\begin{equation}
U_{AD} = - \langle \phi_0 | V_{np}(U_{nA} + U_{pA}) |\phi_0 \rangle,
\end{equation}
with $\phi_0$ being the first Weinberg eigenstate. A detailed discussion of the advantages of ADWA can be found in \cite{Nguyen2010}. 

Note that the beam energies used for the transfer reactions studied are consistently chosen to match the sum of the neutron and proton energies in the incoming  channel. ADWA transfer angular distributions are obtained with the reaction code \textsc{nlat} \cite{nlat}.

\section{Results}
\label{results}

\IBL{As stated in the introduction, the goal of this work is to study the uncertainties in the optical potential through modern UQ tools, and explore experimental conditions that may lead to reducing the uncertainties on the resulting observables. Our UQ methods are general:  their applicability is valid as long at the optical potential concept is a good approximation. We thus choose as targets two doubly magic nuclei in  different regions of the nuclear chart, namely $^{48}$Ca and $^{208}$Pb. Nucleon elastic scattering off of these targets can be well described by the optical model. Because we are also interested in propagating the uncertainties to (d,p) reactions, our applications  include beam energies in the range of 10-65 MeV.}

We first consider neutron and proton elastic scattering on $^{48}$Ca at 12 MeV and on $^{208}$Pb at 30 MeV. 
We also include in our study proton elastic scattering on these targets are 21 MeV and 61 MeV respectively (corresponding to the exit channel energies for the (d,p) process). Applying the Bayesian procedure described in Section \ref{theory}, we obtain parameter posterior distributions for the optical potentials and uncertainty intervals for the elastic scattering angular distributions.
The corresponding parameter posteriors are then used to propagate the uncertainty to the transfer $^{48}$Ca(d,p)$^{49}$Ca(g.s.) at $E_d=21$ MeV and $^{208}$Pb(d,p)$^{208}$Pb(g.s.) at $E_d=61$ MeV.  Here, we explore four different experimental conditions in an attempt to reduce the uncertainties in the calculated angular distributions.

\subsection{Angular coverage}

Because all of our previous studies \cite{Lovell2015,Lovell2017,Lovell2018,King2018,King2019} use angular distributions as our external information, $p(D)$, we first explore the information content in the various ranges of angles to determine whether, by varying the angular grid, one can reduce the uncertainties. The angular distributions of the cross sections can be expressed by a partial wave decomposition:
\begin{equation}
\frac{d\sigma}{d\Omega} = \frac{1}{4k^2}\left\vert \sum_{L=0}^{\infty}(2L+1)P_L(\cos\theta)(S_L-1) \right\vert^2,
\end{equation}
where $k$ is the incoming momentum in the center of mass, $P_L(\cos\theta)$ are the Legendre Polynomials, and $S_L$ are the S-matrix elements \cite{ReactionsBook}. From this relationship one can expect that constraining one angle provides constraints to other angles (correlations between angles have been discussed in \cite{Lovell2017}). However, it is also understood that, for proton elastic scattering, the forward angles are dominated by the Coulomb interaction and it is only the more backward angles that contain the desired optical potential information. Finally, for the largest angles corresponding to central collisions, one does not expect the optical model to hold and  therefore one may obtain disparate results. This qualitative analysis can now be rigorously quantified with Bayesian methods.

We start by carefully choosing an angular grid that is well suited to each particular case: we take 6-12 data points between each pair of minima in the angular distribution. This level of discretization provides a statistically significant set and is comparable to the number of data points present in typical experimental data sets. We then use the Koning-Delaroche optical potential \cite{kd2003} to generate our cross section reference data set, including angles from $\theta=5-160^\circ$. The Bayesian procedure is then run using this data, and we refer to the results of this calculation as {\it full}. We also generate a second set of cross section data by dropping all angles backwards of $\theta=100^\circ$ (we refer to this calculation as {\it forward}).  Finally, we use the original data set from $\theta=5-160^\circ$ degrees and drop every other angle, reducing the number of data points by half (we refer to this calculation as {\it reduced}). 

\begin{figure}[t]
\begin{center}
\includegraphics[width=0.5\textwidth]{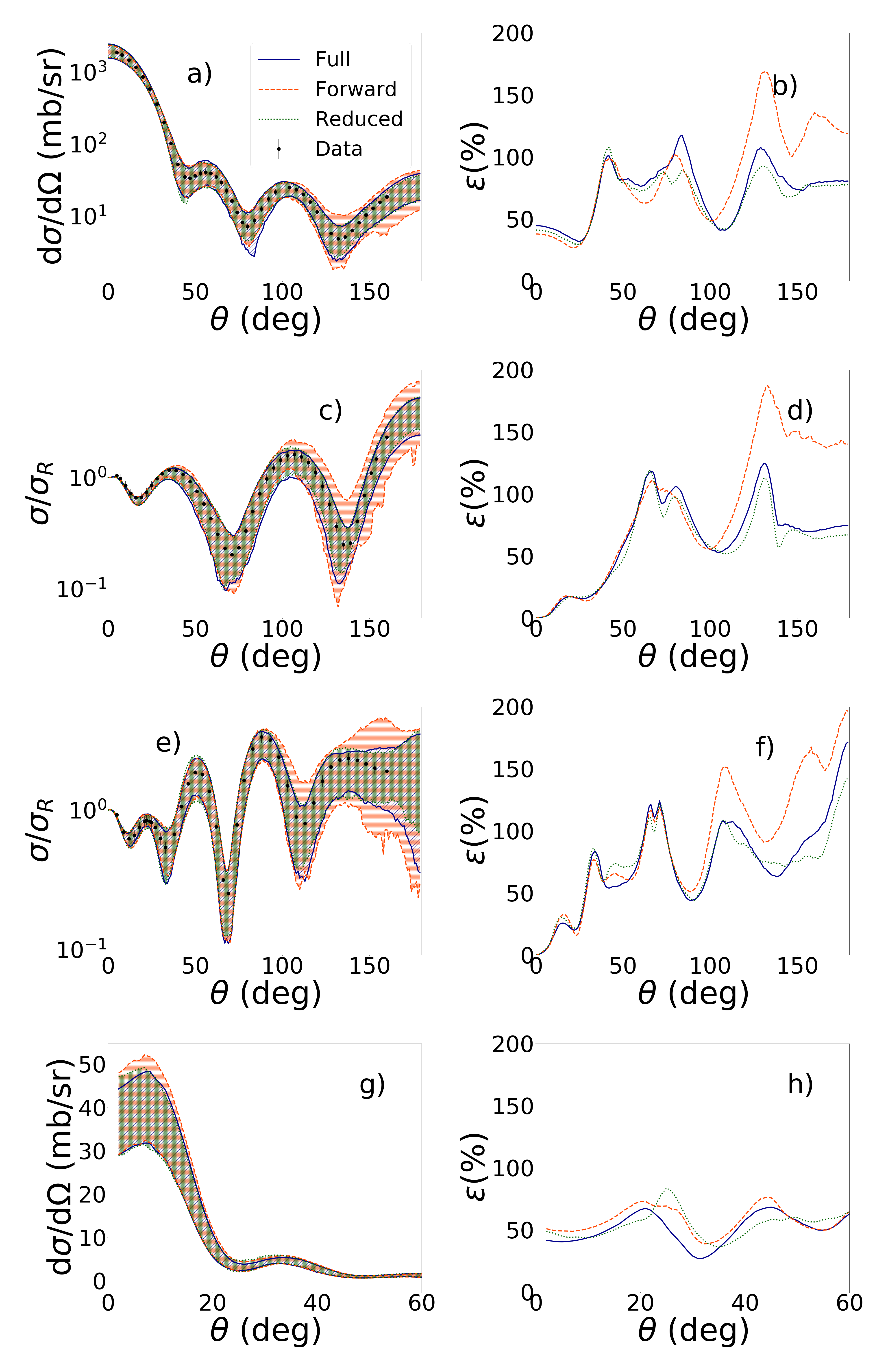}
\end{center}
\caption{A comparison of results using the full angular range (blue solid line)  with those where only forward angles are used (orange dashed) or half of the data points are considered (green dotted): a) and b)  $^{48}$Ca(n,n) at 12 MeV 95$\%$ confidence intervals and percentage uncertainty plot;
 c) and d)  $^{48}$Ca(p,p) at 12 MeV 95$\%$ confidence intervals and percentage uncertainty plot; 
 e) and f)  $^{48}$Ca(p,p) at 21 MeV 95$\%$ confidence intervals and percentage uncertainty plot;
 and g) and h) $^{48}$Ca(d,p) at 21 MeV 95$\%$ confidence bands and percentage uncertainty plot.}
 \label{fig-ang-ca}
\end{figure}

\begin{figure}[t]
\begin{center}
\includegraphics[width=0.5\textwidth]{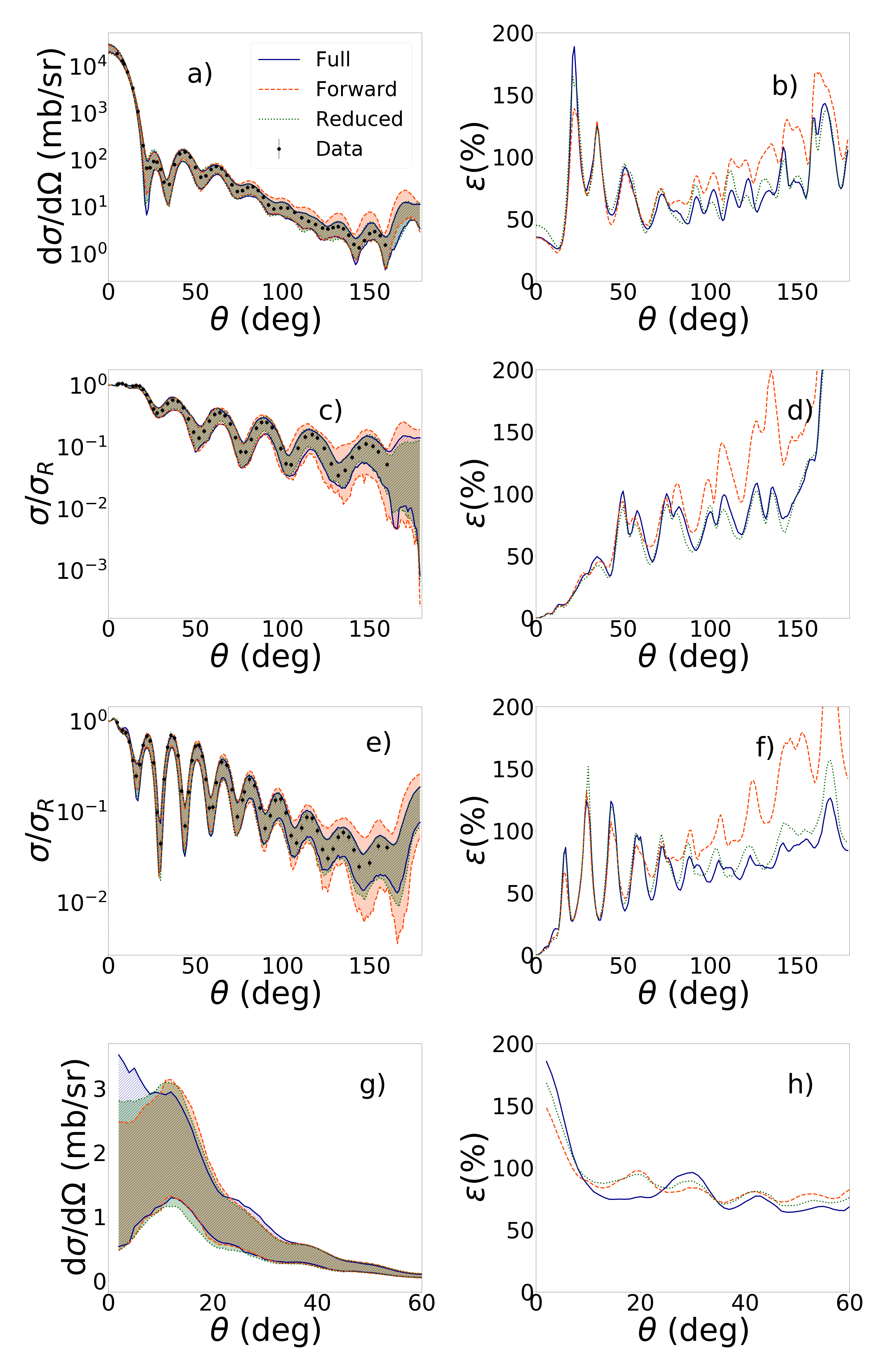}
\end{center}
\caption{A comparison of results using the full angular range (blue solid line)  with those where only forward angles are used (orange dashed) or half of the data points are considered (green dotted): a) and b)  $^{208}$Pb(n,n) at 30 MeV 95$\%$ confidence intervals and percentage uncertainty plot;
 c) and d)  $^{208}$Pb(p,p) at 30 MeV 95$\%$ confidence intervals and percentage uncertainty plot; 
 e) and f)  $^{208}$Pb(p,p) at 61 MeV 95$\%$ confidence intervals and percentage uncertainty plot;
 and g) and h) $^{208}$Pb(d,p) at 61 MeV 95$\%$ confidence intervals and percentage uncertainty plot.}
 \label{fig-ang-pb}
\end{figure}

Fig. \ref{fig-ang-ca}  contains all these results for $^{48}$Ca: panels a) and c) show the 95\% confidence intervals for the elastic scattering of neutrons and protons off $^{48}$Ca at 12 MeV; panels b) and d) are the corresponding percentage error, quantified as the width of the 95\% confidence interval divided by the mean, multiplied by 100;  panels e) and f)  show the same quantities but now for proton scattering off $^{48}$Ca at 21 MeV; and finally, panels g) and h) are the predicted quantities for $^{48}$Ca(d,p)$^{49}$Ca at $E_d=21$ MeV. The identical quantities for  $^{208}$Pb are shown in Fig. \ref{fig-ang-pb}. Shown are the results with the {\it full} angular range (blue solid), with {\it forward} angular range (orange dashed) and with {\it reduced} the data points (green dotted). 

For all cases, the uncertainties obtained with the {\it reduced} data sets are very similar to those obtained when the {\it full} set is considered (the parameter posteriors, not shown here, are also overlapping). We repeated this analysis and included a larger set of angles (considering data points for every degree) and still the results were identical. As far as elastic scattering is concerned, adding more angles to the angular distribution offers no additional constraints to the optical potential parameters. Data that provides only a rough outline of the diffraction pattern is sufficient. This is consistent with the observation that different angles are correlated \cite{Lovell2017}. 

Concerning the results when dropping the larger angles, we see only small changes in the posterior distributions for some parameters of the imaginary terms in the optical potential. The uncertainty on the cross sections does increase at backward angles for the elastic distribution, and in some cases the uncertainty at the intermediate angles is reduced.  We attribute this complex picture to the fact that, at the most backward angles, the optical model does not provide a reliable description of the process and there is a chance of ending up with artificial parameters. 

When these results are propagated to the transfer, the effects are mixed due to the non linear form by which the various parameter posteriors enter in the calculation. For example, if we focus only on the forward angles, dropping the larger angles in the fit produces a small increase in the uncertainty for the transfer cross section on $^{48}$Ca but a significant reduction of the uncertainty for the transfer cross section on $^{208}$Pb.

\subsection{Nearby energy range} 

It has been argued that local optical potentials can be better constrained by fitting data taken at several nearby energies (e.g. \cite{bassel1964}). The assumption is that by using a small range of energies around the energy of interest, one reduces possible spurious effects and imposes a tighter constraint. We have explored this idea by generating another set of data at a nearby energy  and comparing the confidence interval obtained when only a single set $\sigma^{E_1}_{el}(\theta)$ is included in the method, to the case when both the original and the additional set $\sigma^{E_2}_{el}(\theta)$ of mock data are included. As  additional sets, we included mock data generated with \cite{kd2003} at 14 MeV for n$+^{48}$Ca, 14 MeV for p$+^{48}$Ca, 24 MeV for p$+^{48}$Ca, 32 MeV for n$+^{208}$Pb, 32 MeV for p$+^{208}$Pb, and 65 MeV for p$+^{208}$Pb.
There are two possibilities of incorporating the respective pair of angular distributions. In the first, we find a joint parameterization for the two data sets (this approach is denoted by {\it multiple}). We take equal weights in the $\chi^2$-function for the two data sets and consider the cross sections at the exact same angles. In a second procedure,  we first consider the nearby energy data set and generate the posteriors from it. These are then introduced as priors in the fit to the first data set (this approach is denoted by {\it sequential}).

\begin{figure}[t]
\begin{center}
\includegraphics[width=0.5\textwidth]{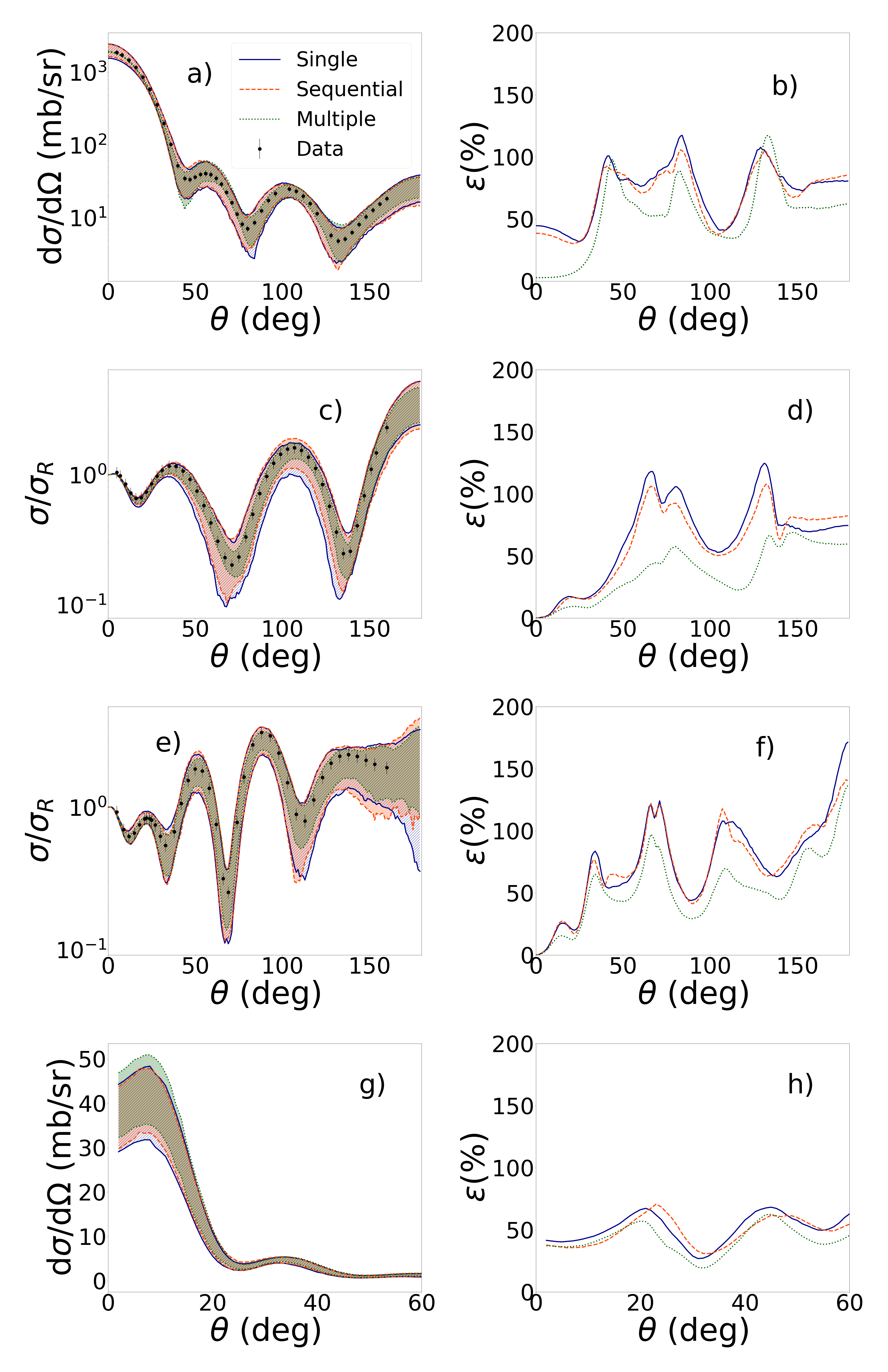}
\end{center}
\caption{A comparison of results using a single data set (blue solid line)  with those using two sets at nearby beam energies, either sequentially (green dotted) and simultaneously (orange dashed): a) and b)  $^{48}$Ca(n,n) at 12 MeV 95$\%$ confidence intervals and percentage uncertainty plot;
 c) and d)  $^{48}$Ca(p,p) at 12 MeV 95$\%$ confidence intervals and percentage uncertainty plot; 
 e) and f)  $^{48}$Ca(p,p) at 21 MeV 95$\%$ confidence intervals and percentage uncertainty plot;
 and g) and h) $^{48}$Ca(d,p) at 21 MeV 95$\%$ confidence intervals and percentage uncertainty plot.}
 \label{fig-en-ca}
\end{figure}

\begin{figure}[t]
\begin{center}
\includegraphics[width=0.5\textwidth]{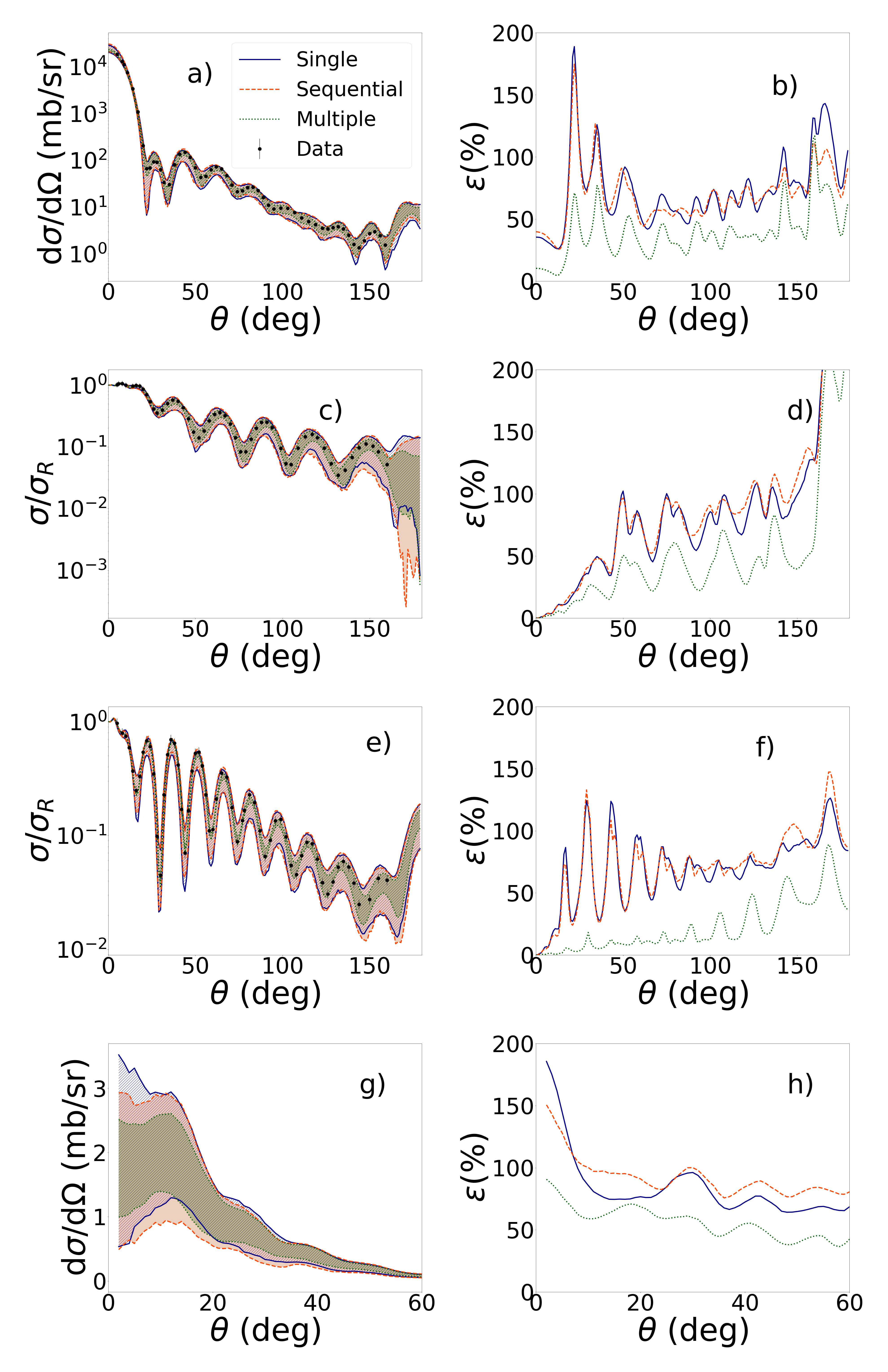}
\end{center}
\caption{A comparison of results using a single data set (blue solid line)  with those using two sets and nearby beam energies, either sequentially (green dotted) and simultaneously (orange dashed): a) and b)  $^{208}$Pb(n,n) at 30 MeV 95$\%$ confidence intervals and percentage uncertainty plot;
 c) and d)  $^{208}$Pb(p,p) at 30 MeV 95$\%$ confidence intervals and percentage uncertainty plot; 
 e) and f)  $^{208}$Pb(p,p) at 61 MeV 95$\%$ confidence intervals and percentage uncertainty plot;
 and g) and h) $^{208}$Pb(d,p) at 61 MeV 95$\%$ confidence intervals and percentage uncertainty plot.}
 \label{fig-en-pb}
\end{figure}

Fig.\ref{fig-en-ca} (Fig.\ref{fig-en-pb}) show the results obtained for the 95\% confidence intervals for the reactions on $^{48}$Ca ($^{208}$Pb): we show on the left the differential angular distributions as a function of scattering angle and on the right the uncertainty interval as a percentage. We note that if the two data sets are included sequentially, no improvement is found, although the sequential method finds a slightly different  minimum in parameter space. On the contrary, when both sets are included simultaneously, the minimum found in parameter space is the same as that obtained when using only the single original data set, but now, with the additional constraint, the uncertainty can be further reduced.
For $^{48}$Ca, the inclusion of the additional data at a nearby energy produces at most a modest reduction on the uncertainty intervals in  elastic scattering or transfer, except for neutron elastic scattering where the effect is important.
For $^{208}$Pb the improvement is very significant for all elastic scattering cases considered and results in a factor of 2 reduction in the uncertainty interval at forward angles for the transfer cross section.

We have verified that there are two factors contributing to the different findings in $^{48}$Ca and $^{208}$Pb. The first has to do with the different energy regimes. If one repeats the process for $^{48}$Ca reaction at the same energies as those for $^{208}$Pb, the resulting uncertainties are significantly reduced, in line with the results show in Fig.\ref{fig-en-pb}. The second has to do with how close the nearby energy is to the original energy. This value cannot be too far from the original value, otherwise the energy dependence of the optical potential would need to be explicitly considered (introducing a larger array of parameters). However, this nearby energy should not be too close to the original, otherwise the added data becomes redundant. If we repeat the $^{208}$Pb calculation using as the second energy $E=35$ instead of $E=32$ MeV (same percent difference as in the $^{48}$Ca case), we obtain only a modest improvement on the uncertainty, a result more in line with Fig. \ref{fig-en-ca}.


\subsection{Experimental error bar}

One obvious way to impose a more stringent constraint is by reducing the error bars on the experimental data. Up to this point, we have considered as our standard value the nominal error on all data of 10\%. While this value is more common in stable beam experiments where statistics are plentiful (in some cases the error obtained in stable beam experiments is even lower than 5\%), the same cannot be said for radioactive beam experiments in inverse kinematics. In those cases, statistics often limit errors closer to 20\%.  In this subsection, we explore the consequences on the confidence intervals predicted by theory of reducing the experimental error bar from the nominal 10\% to 5\% or increasing it to 20\%.

Our results are summarized in Table \ref{tab-error}. We consider the percentage error obtained when the data has an error of 20\%, averaged over angle ($\varepsilon_{20}$) and 
the percentage error obtained when the data has an error of 10\%, averaged over angle ($\varepsilon_{10}$). The ratio  $\Delta \varepsilon _{20/10}=\frac{\varepsilon_{20}}{\varepsilon_{10}}$
corresponds to the second column in Table \ref{tab-error}. The third column shows the ratio between the results assuming 10\% error on the data, and 5\% error on the data:
$\Delta \varepsilon _{10/5}=\frac{\varepsilon_{10}}{\varepsilon_{5}}$. 

Expectedly, reducing/increasing the error bar on the elastic scattering data does translate to a reduction/increase in the predicted uncertainty. However this effect is  not  directly proportional to the change in error, underlining the non-linearity of the effect.
The conclusion here is that while there is significant gain by reducing the error bar for  reactions on both $^{48}$Ca and on $^{208}$Pb,  the magnitude of the improvement depends on the particular reaction and the angular region considered and is seldom a factor of 2. This aspect may make it less attractive to the experimental community to work on increased precision in the experiment.

\begin{table}[t]
\begin{center}
\begin{tabular}{|c|r|r|r|}
\hline Reaction&$\Delta \varepsilon _{20/10}$ &$\Delta \varepsilon _{10/5}$  \\ \hline
$^{48}$Ca(n,n) at 12 MeV&1.53 &1.94	
  \\ \hline
$^{48}$Ca(p,p) at 12 MeV&1.68 &1.71	
  \\ \hline
$^{48}$Ca(p,p) at 21 MeV &1.55 &1.74	
  \\ \hline
  $^{48}$Ca(d,p) at 21 MeV &1.68 &1.52	
  \\ \hline
  $^{208}$Pb(n,n) at 30 MeV&1.62 &1.79
  \\ \hline
  $^{208}$Pb(p,p) at 30 MeV &1.39 &1.61
  \\ \hline
  $^{208}$Pb(p,p) at 61 MeV &1.99 &1.74
  \\ \hline
  $^{208}$Pb(d,p) at 61 MeV &1.41 &1.58
  \\ \hline
\end{tabular}
\caption{Ratio of the average uncertainties obtained with changing the experimental error bars on the elastic scattering data. More details in the text.}
\label{tab-error}
\end{center}
\end{table}

\subsection{Additional reaction data}

Other groups have found success in including more information in the fitting procedure, particularly with total reaction cross sections \cite{Atkinson2018}. Motivated by that work, we investigated the impact on our uncertainties when including the total (reaction) cross section data in addition to differential cross section data for neutron (proton) elastic scattering. Again, we use the optical model parameterization of Ref. \cite{kd2003} to generate these results and apply equal weight to this  additional data point as that of the whole differential angular distribution used previously. This choice is rather ambiguous, but by doing it this way, we maximize the effects of this additional information.

\begin{figure}[t]
\begin{center}
\includegraphics[width=0.5\textwidth]{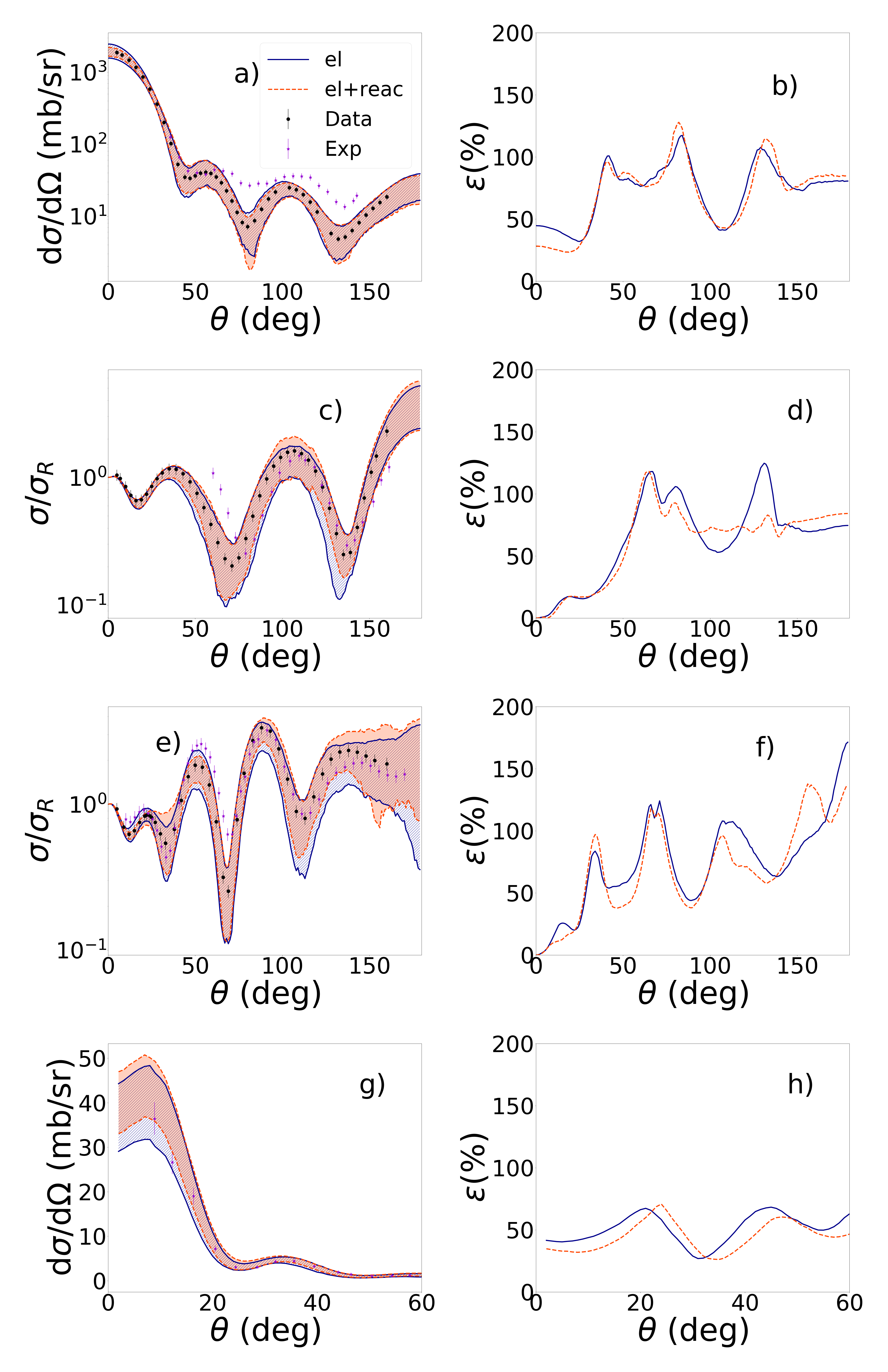}
\end{center}
\caption{A comparison of results using elastic scattering angular distributions (blue solid line)  with those using in addition the total (reaction) cross section  (orange dashed): a) and b)  $^{48}$Ca(n,n) at 12 MeV 95$\%$ confidence intervals and percentage uncertainty plot;
 c) and d)  $^{48}$Ca(p,p) at 12 MeV 95$\%$ confidence intervals and percentage uncertainty plot; 
 e) and f)  $^{48}$Ca(p,p) at 21 MeV 95$\%$ confidence intervals and percentage uncertainty plot;
 and g) and h) $^{48}$Ca(d,p) at 21 MeV 95$\%$ confidence intervals and percentage uncerainty plot. \IBL{Mock data from KD (black circles) and real experimental data (grey  stars) are also shown.}}
 \label{fig-xs-ca}
\end{figure}

\begin{figure}[t]
\begin{center}
\includegraphics[width=0.5\textwidth]{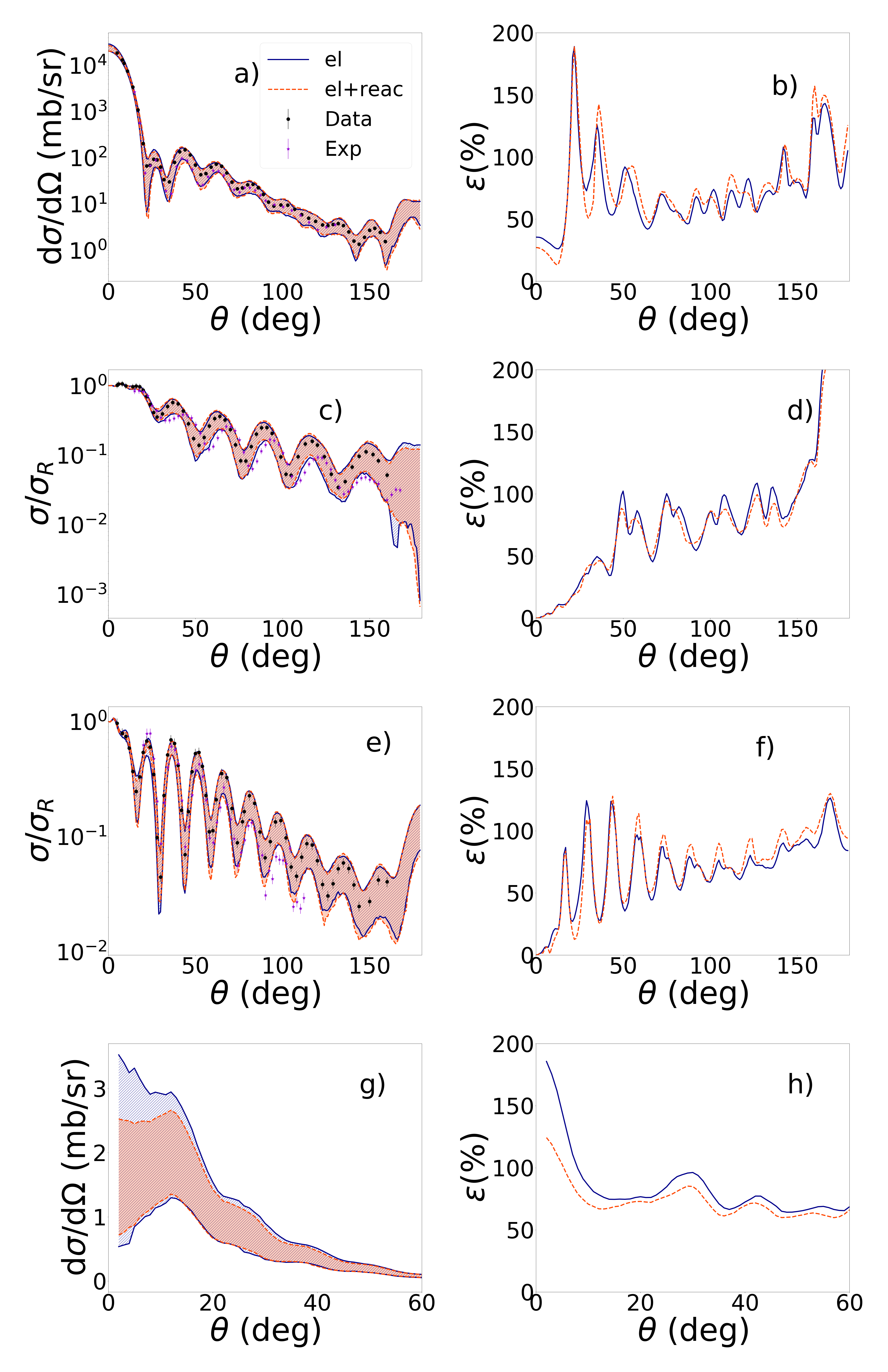}
\end{center}
\caption{A comparison of results using elastic scattering angular distributions (blue solid line)  with those using in addition the total (reaction) cross section (orange dashed): a) and b)  $^{208}$Pb(n,n) at 30 MeV 95$\%$ confidence intervals and percentage uncertainty plot;
 c) and d)  $^{208}$Pb(p,p) at 30 MeV 95$\%$ confidence intervals and percentage uncertainty plot; 
 e) and f)  $^{208}$Pb(p,p) at 61 MeV 95$\%$ confidence intervals and percentage uncertainty plot;
 and g) and h) $^{208}$Pb(d,p) at 61 MeV 95$\%$ confidence intervals and percentage uncertainty plot. \IBL{Mock data from KD  (black circles) and real experimental data (grey  stars) are also shown.}}
 \label{fig-xs-pb}
\end{figure}

The results including the reaction data (orange dashed lines) are compared to those including only elastic angular distribution (blue solid lines) in Figs. \ref{fig-xs-ca} and \ref{fig-xs-pb}.
Including the total cross section (or reaction cross section) can result in a reduction on the uncertainties in the elastic scattering distribution, but the magnitude of the effects depends on the angle and particular reaction considered. For the transfer cross sections, we  find that, for $^{48}$Ca, there is no significant change in the uncertainty intervals, but the contrary is true for $^{208}$Pb, where we find a reduction of $\approx 20$\% at forward angles. This can be explained by the fact that the optical potential posteriors generated when including the total (reaction) cross sections shift significantly for $^{208}$Pb, particularly in the imaginary depths and diffusenesses. Although the parameter posteriors themselves are not narrower, they result in a narrower range for this observable. The differences between the $^{48}$Ca and the $^{208}$Pb cases are primarily due to the different energy regimes. When we repeated the $^{48}$Ca calculations at the same energies as the $^{208}$Pb, we obtained similar reductions in the uncertainty intervals when including reaction cross sections. 
\IBL{For completeness, we present in Table \ref{tab-xs} the predicted 95\% intervals for the total (reaction) cross section $\sigma^{Bayes}_{range}$ from this work (when both elastic and total (reaction) cross sections are included in the likelihood). The predicted intervals are consistent with the mock data $\sigma^{KD}$. }

\begin{table}[H]
\begin{center}
\begin{tabular}{|c|r|r|r|r|r|}
\hline reaction & E (MeV) & $\sigma^{CI}$ (mb) & $\sigma^{KD}$ (mb)   \\ \hline
$^{48}$Ca(n,n) &12 &1221--1436  &1322  \\ \hline
$^{48}$Ca(p,p) & 12 &920--1095  &999     \\ \hline
$^{48}$Ca(p,p) & 21 &984 --1189 &1083  \\ \hline
$^{208}$Pb(n,n) & 30 &2324 -- 2688   & 2486 \\ \hline
$^{208}$Pb(p,p) & 30 &1688 -- 2191 & 1891   \\ \hline
$^{208}$Pb(p,p) & 61 &2037 -- 2254 & 2133 \\ \hline
\end{tabular}
\caption{Comparing Bayesian interval for total (reaction) cross sections with mock data.}
\label{tab-xs}
\end{center}
\end{table}

It is also known that polarization observables can be used to further constrain the optical potential. As a first step in exploring the information content of polarization observables, using the same global optical potential \cite{kd2003}, we have generated  vector analyzing powers,  $Re(iT_{11})=\sqrt{3}/2 \; A_y$, for all the elastic-scattering cases in our study. We have introduced the nominal uncertainty on the data of $\epsilon=|10\%iT_{11}|$, just as before. We first apply the Bayesian procedure to constrain the same 9 optical potential parameters as in the previous sections but now with this polarization data alone. The resulting posteriors were much narrower (by an order of magnitude).  This result is unrealistic and is caused by the artificial error bars:  the $Re(iT_{11})$ angular distribution oscillates around zero, and the percent error ends up introducing absolute errors close to zero, and driving the minimization procedure.  We thus corrected this by taking a minimum error representing a lower bound from systematic uncertainties in the measurement: when $|Re(iT_{11})|$ becomes lower than 5\% of it's maximum value, we take $\epsilon=|5\%Max(Re(iT_{11}))|$. This choice is rather ambiguous and the results obtained are more in line with the uncertainty intervals produced in the elastic-scattering angular distributions shown in Fig. \ref{fig-xs-ca} and \ref{fig-xs-pb}.
\IBL{Further work is in progress to incorporate polarization consistently and correctly in the definition of the likelihood and will be reported elsewhere. }

\subsection{Confronting our results with real data}

\IBL{As mentioned before, in order to have control over the experimental conditions, we used mock data generated from the global optical model \cite{kd2003}. Global parameterizations such as \cite{kd2003} cannot provide a perfect reproduction for  elastic scattering of any single data set. However, on average, these parameterizations should be able to provide a fair description of reality. Most importantly, we have now quantified the uncertainty in the determination of the optical model parameters and therefore it is useful to confront the predicted uncertainties with real data. }

\IBL{In Fig. 5 we have included real data (open grey stars), with real error bars,  from Refs. \cite{data-ca48n12,data-ca48p12,data-ca48p21} for scattering on $^{48}$Ca. Data from Refs. \cite{data-pb208n30,data-pb208p30,data-pb208p61} for scattering on $^{208}$Pb is used in Fig. 6. The percentage of time that the real data (with real error bars) falls into the predicted uncertainty interval varies from case to case, but ranges from $75-100$\%. These values are to be compared with the 95\% confidence level calculated. The exception is for n+$^{48}$Ca at 12 MeV, for which the empirical coverage is only $26$\%. Clearly for this case the KD potential \cite{kd2003} does not provide a good description. In addition, we have also included (d,p) data from \cite{metz1975}  in Figs. 5g (no additional normalization is applied).  We conclude that the UQ intervals obtained with mock data are physically reasonable.}

\section{Conclusions}
\label{conclusions}

In this work, we use the Markov-Chain Monte Carlo Bayesian approach to explore different aspects of experimental conditions in the attempt to reduce the uncertainties associated with elastic scattering and transfer reactions. We perform systematic studies of neutron and proton elastic scattering on $^{48}$Ca and $^{208}$Pb and the associated  (d,p) reactions using the three-body model ADWA for the reaction. We use mock data generated from a global optical potential so there is total control on the assumed conditions for the experiment.

As a first step, we explore the information content of the angular distribution. We compare uncertainty intervals obtained for elastic and transfer observables using a dense angular grid, a sparse angular grid, and including only forward angles. We find the results are not significantly sensitive to the number of angles included, as long as it can roughly capture the diffraction pattern. Expectedly, by not including the backward angles in the fit, the uncertainties for the elastic-scattering angular distributions increase significantly at backward angles, but this does not translate into a larger uncertainty in the transfer, a result that points to the non-linear nature of the problem.

Second, we explore the constrain coming from an additional angular distribution data set measured at a nearby energy. For this case, we demonstrate that when including two sets of data at nearby energies simultaneously in the procedure, one can improve the uncertainty intervals by up to a factor of 2. However, for the two cases considered here, the method works best at higher energies and if the nearby energies are chosen to be $\sim7$\% apart.

We next explored the impact of the uncertainty intervals coming from reducing the experimental error bar. While the uncertainty decreases with smaller errors bars as one would expect, the gain is not directly proportional to the reduction factor for the error bars. In most cases, there is a loss coming from the complex way in which the various parameter posteriors work together to produce the desired observable. 

Finally, we considered the inclusion of the total (reaction) cross sections in the Bayesian procedure. Although the results depend on the case considered, we find that the inclusion of total (reaction) cross section can offer a reduction on the uncertainty in the elastic and transfer observables, but the magnitude depends on the particular reaction, beam energy and the angular range. Finally, we also performed preliminary work to include vector analysing powers in the Bayesian procedure. Our results indicate that a dedicated study, exploring other statistical tools, is needed to obtain useful results.  Such a study is in progress.

\IBL{The UQ tools we have so far developed make use of global phenomenological potentials and  the adiabatic wave approximation for (d,p) reactions. However the UQ tools themselves are very general and can be coupled with more advanced optical potential approaches and  reaction theories. As the UQ tools become established, future planned collaborations entail attaching the UQ framework to upgraded optical model approaches and reaction theories for the most reliable interpretation of the physics.}

\IBL{As mentioned in the introduction, reactions at the limits of stability are particularly relevant for astrophysics because a fraction of the production of heavy nuclei involves neutron capture reactions on unstable nuclei. Since we cannot measure neutron capture on exotic nuclei directly, (d,p)  transfer reactions are used as an indirect probe \cite{bardayan2016}.  These transfer reactions are typically performed in inverse kinematics, where the unstable nucleus is the beam and the target is the light particle (either the proton or the deuteron). Currently, with the rapid development of a number of active target time projection chambers (AT-TPC), and given their astonishing tracking capabilities, we are already able to scan energy in one single experiment due to the beam energy loss in the active target. This is very fortunate given that, of the cases explored in this work, we find that the most promising case for reducing the uncertainties in the optical potential is by including data at nearby energies. }

\IBL{Given the demand for beam time and the limited resources, we expect data on rare isotopes will continue to be scarce. This makes it even more important that we know the theory uncertainties associated with interpreting any given measurement involving reactions with unstable beams and using modern UQ tools to help identify the experimental design that will hold the most information. }


\begin{acknowledgments}
This work was supported by the National Science
Foundation under Grant  PHY-1811815 and was performed under the auspice of the U.S. Department of Energy by Los Alamos National Laboratory under Contract 89233218CNA000001.  We gratefully acknowledge the support of the U.S. Department of Energy through the LANL/LDRD Program and the Center for Non Linear Studies. This work relied on iCER and the High Performance Computing Center at Michigan State University for computational resources. 
\end{acknowledgments}

\bibliography{bayes-exp-gk} 

\end{document}